\documentclass[showpacs,preprintnumbers,pra,twocolumn,10pt]{revtex4}
\usepackage{mathrsfs}
\usepackage{dcolumn}
\usepackage{bm}
\usepackage{graphicx}
\usepackage{amsfonts}
\usepackage{amssymb}
\usepackage{amsmath}

\begin{document}

\title{Proposed method for direct measurement of non-Markovian character of
the qubits coupled to bosonic reservoirs}
\author{Z. Y. Xu$^{1,2}$}
\author{W. L. Yang$^{1}$}
\author{M. Feng$^{1}$}
\altaffiliation{mangfeng@wipm.ac.cn}
\affiliation{$^{1}$State Key Laboratory of Magnetic Resonance and Atomic and Molecular
Physics, Wuhan Institute of Physics and Mathematics, Chinese Academy of
Sciences, Wuhan 430071, China \\
$^{2}$Graduate School of the Chinese Academy of Sciences, Beijing 100049,
China}

\begin{abstract}
The non-Markovianity is a recently proposed characterization of the
non-Markovian behavior in an open quantum system, based on which we first
present a practical idea for directly measuring the non-Markovian character
of a single qubit coupled to a zero-temperature bosonic reservoir, and then
extend to investigate the dynamics of two noninteracting qubits subject to
two reservoirs respectively with a lower bound of non-Markovianity. Our
scheme, with no need of optimization procedures and quantum state
tomography, is helpful for experimental implementation.
\end{abstract}

\maketitle

Realistic quantum systems are fragile due to unavoidable interaction with
the environment \cite{open book,open book2}. For this reason, the dynamics
of open quantum systems have attracted much attention in the investigations
of modern quantum theory, particularly of quantum information processing
(QIP) \cite{QIP Book}. Over past decades, the conventionally employed
Markovian approximation with the assumption of infinitely short correlation
time of the environment has experienced more and more challenges due to
advance of experimental techniques \cite{open book}, and non-Markovian
features have been observed in some physical systems, e.g., atomic and
molecular systems \cite{exp-molecular}, high-Q cavity systems \cite%
{exp-High-Q} and solid state systems \cite{exp-solid}. To understand these
non-Markovian effects, there have been various kinds of analytical and
numerical methods developed so far \cite{open book,quantum
trajectory,pseudomode,NMQJ,semi-MP}, such as non-Markovian quantum
trajectories \cite{quantum trajectory}, pseudomodes \cite{pseudomode},
non-Markovian quantum jumps \cite{NMQJ}, and quantum semi-Markov processes
\cite{semi-MP}.

Among the recent investigations for non-Markovian behavior \cite%
{M,non-Markovianity,nonM2,nonM3}, a particular approach using
non-Markovianity contains the exact characterization of the non-Markovian
behavior of a quantum process without making any approximation for the
dynamics \cite{non-Markovianity}. The non-Markovianity is defined to
quantify the total amount of information flowing from the environment back
to the system through a quantum process $\Phi (t)$ with $\rho (t)=\Phi
(t)\rho (0),$ where $\rho (0)$ and $\rho (t)$ are the density operators of
the system at initial time and at arbitrary time. In Ref. \cite%
{non-Markovianity}, the information is characterized by the trace distance $%
\mathcal{D}\left( \rho _{1},\rho _{2}\right) =\frac{1}{2}tr\left\vert \rho
_{1}-\rho _{2}\right\vert $ of two quantum states $\rho _{1}$ and $\rho _{2},
$ describing the distinguishability between the two states, and satisfying $%
0\leq \mathcal{D}\leq 1$ \cite{QIP Book}$.$ $\mathcal{D}$ maximally reaches
1 when the two states are totally distinguishable and approaches 0 for two
identical states \cite{QIP Book}. The direction of information flow is
dependent on the slope of $\mathcal{D}\left( \rho _{1}(t),\rho
_{2}(t)\right) ,$ i.e., when $\partial _{t}\mathcal{D}\left( \rho
_{1}(t),\rho _{2}(t)\right) <0,$ the information dissipates to the
environment and vice versa. Therefore, the non-Markovianity could be
calculated by
\begin{eqnarray}
\mathcal{N}(\Phi ) &=&\underset{\rho _{1,2}(0)}{\max }\sum_{n}[\mathcal{D}%
\left( \rho _{1}\left( \tau _{n}^{\max }\right) ,\rho _{2}\left( \tau
_{n}^{\max }\right) \right)   \notag \\
&&-\mathcal{D}(\rho _{1}(\tau _{n}^{\min }),\rho _{2}(\tau _{n}^{\min }))],
\end{eqnarray}%
with $\tau _{n}^{\min }$ ($\tau _{n}^{\max }$) the time point when $\mathcal{%
D}\left( \rho _{1}(t),\rho _{2}(t)\right) $ reaches the $n$th local minimum
(maximum). Eq. (1) can be carried out by summing up the amount of the
increase of the trace distance over each time interval [$\tau _{n}^{\min
},\tau _{n}^{\max }$] for any pair of initial states $\rho _{1}(0)$ and $%
\rho _{2}(0)$, where the maximum is considered as the non-Markovianity $%
\mathcal{N}(\Phi )$. Since this is a problem of optimization, however, we
have to consider all pairs of initial states in our calculation, which is
inconvenient and impractical, especially from the viewpoint of experimental
exploration.

In this work, we show that the optimization problem of
$\mathcal{N}(\Phi )$ could be simplified to an effectively
computable expression in the case of a single qubit coupled to a
zero-temperature bosonic reservoir. Moreover, for two independent
qubits coupled to two bosonic reservoirs respectively, we
investigate a lower bound of the non-Markovianity. The favorable
feature of our method is the possibility to connect the
non-Markovianity to the population of the qubit in the excited
state, which could be detected directly in experiments without the
requirement of tomographic reconstruction of the density matrix. In
addition, our result also makes it possible to have an easy
evaluation of non-Markovianity even without much information about
the interaction between the qubit and the reservoir.

We first consider a single qubit coupled to a zero-temperature bosonic
reservoir. The Hamiltonian in units of $\hbar =1$ is given by
\begin{equation}
H=\omega _{0}|e\rangle \left\langle e\right\vert +\sum_{l}\omega
_{l}a_{l}^{\dagger }a_{l}+\sum_{l}\left( g_{l}|e\rangle \left\langle
g\right\vert a_{l}+g_{l}^{\ast }|g\rangle \left\langle e\right\vert
a_{l}^{\dagger }\right),
\end{equation}
where $\omega _{0}$ is the resonant transition frequency of the qubit
between the excited state $\left\vert e\right\rangle $ and the ground state $%
\left\vert g\right\rangle $. $\omega _{l}$ and $a_{l}$ $(a_{l}^{\dagger })$
are, respectively, the frequency and the annihilation (creation) operator of
the $l$th mode of the reservoir with the coupling constant $g_{l}$ to the
qubit. The dynamics of the single qubit can be represented by the reduced
density matrix \cite{open book}
\begin{equation}
\rho ^{S}\left( t\right) =\left(
\begin{array}{ll}
\rho _{ee}^{S}\left( 0\right) \left\vert b(t)\right\vert ^{2} & \rho
_{eg}^{S}\left( 0\right) b(t) \\
\rho _{ge}^{S}\left( 0\right) b^{\ast }(t) & 1-\rho _{ee}^{S}\left( 0\right)
\left\vert b(t)\right\vert ^{2}%
\end{array}
\right)
\end{equation}
in the qubit basis $\left\{ \left\vert e\right\rangle ,\text{ }\left\vert
g\right\rangle \right\},$ where the superscript $S$ of $\rho $ represents
the single-qubit case. $b(t)$ can be interpreted as the amplitude of the
upper level $\left\vert e\right\rangle $ of the qubit initially prepared
with $\rho _{ee}^{S}\left( 0\right) =1$ and $b\left( 0\right) =1,$ and is
given by the inverse Laplace transform
\begin{equation}
b(t)=\mathscr{L}^{-1}\left[ \frac{1}{s+F(s)}\right] ,
\end{equation}
where the parameter $s$ is a complex number and $F(s)=\mathscr{L}[f(t)]%
=\int_{0}^{\infty }f(t)\exp (-st)dt$ with the correlation function $%
f(t)=\sum_{l}\left\vert g_{l}\right\vert ^{2}e^{i\delta _{l}t}=\int d\omega
J(\omega )e^{i\delta t}$ and $\delta _{(l)}=\omega _{0}-\omega _{(l)}$. The
explicit form of $b(t)$ depends on the specific spectral density of the
reservoir \cite{open book}.

As the density matrix should be of hermiticity, normalization, and
semi-positivity, any pair of initial states could be written as
\begin{eqnarray}
\rho _{1}^{S}(0) &=&\left(
\begin{array}{ll}
\alpha & \beta \\
\beta ^{\ast } & 1-\alpha%
\end{array}%
\right) ,  \notag \\
\rho _{2}^{S}(0) &=&\left(
\begin{array}{ll}
\mu & \nu \\
\nu ^{\ast } & 1-\mu%
\end{array}
\right) ,
\end{eqnarray}
with $\left\vert \beta \right\vert ^{2}\leq \alpha (1-\alpha ),$ $\left\vert
\nu \right\vert ^{2}\leq \mu (1-\mu ),$ ($\beta ,\nu $)$\in \mathcal{\
\mathbb{C}
}$ and $0\leq \alpha $ $(\mu )\leq 1,$ ($\alpha ,\mu $)$\in \mathcal{\
\mathbb{R}
}$. So we have
\begin{eqnarray}
\rho _{1}^{S}(t) &=&\left(
\begin{array}{ll}
\alpha \left\vert b(t)\right\vert ^{2} & \beta b\left( t\right) \\
\beta ^{\ast }b^{\ast }(t) & 1-\alpha \left\vert b(t)\right\vert ^{2}%
\end{array}
\right) ,\text{ }  \notag \\
\rho _{2}^{S}(t) &=&\left(
\begin{array}{ll}
\mu \left\vert b(t)\right\vert ^{2} & \nu b\left( t\right) \\
\nu ^{\ast }b^{\ast }(t) & 1-\mu \left\vert b(t)\right\vert ^{2}%
\end{array}
\right).
\end{eqnarray}
Using the definition of the trace distance, we obtain
\begin{equation}
\mathcal{D}^{S}\left( \rho _{1}^{S}(t),\rho _{2}^{S}(t)\right) =\left\vert
b(t)\right\vert \sqrt{\left\vert b(t)\right\vert ^{2}\left( \alpha -\mu
\right) ^{2}+\left\vert \beta -\nu \right\vert ^{2}},
\end{equation}

In what follows, we consider the case that the qubit interacts resonantly
with a reservoir with Lorentzian spectral distribution
\begin{equation}
J(\omega )=\frac{1}{2\pi }\frac{\gamma _{0}\Gamma ^{2}}{(\omega _{0}-\omega
)^{2}+\Gamma ^{2}},
\end{equation}
with $\gamma _{0}$ the Markovian decay rate and $\Gamma $ the spectral width
of the coupling \cite{open book}, which has been widely employed in quantum
optics \cite{Q-optics}. We may distinguish the Markovian and the
non-Markovian regimes using $\gamma _{0}$ and $\Gamma $: $\gamma _{0}<\Gamma
/2$ means the Markovian regime and $\gamma _{0}>\Gamma /2$ corresponds to
the non-Markovian regime. Therefore, using Eq. (4), we have $b(t)=\exp
(-\Gamma t/2)\left[ \cosh (\kappa t/2)+(\Gamma /\kappa )\sinh (\kappa t/2) %
\right] $, $\gamma _{0}<\Gamma /2$ and $b(t)=\exp (-\Gamma t/2)\left[ \cos
(\kappa t/2)+(\Gamma /\kappa )\sin (\kappa t/2)\right] $, $\gamma
_{0}>\Gamma /2,$ with $\kappa =\sqrt{|\Gamma ^{2}-2\gamma _{0}\Gamma |}$
\cite{open book}. We may check that in the non-Markovian regime, all local
minima of $\left\vert b(t)\right\vert $ approach zeros at $\tau _{n}^{\min
}=2[n\pi -\arctan (\kappa /\Gamma )]/\kappa $ with $n=1,2,3,\cdots$, i.e., $%
\left\vert b(\tau _{n}^{\min })\right\vert =0$. According to Eq. (7), all
nontrivial trace distances own the same monotonicity, so $\mathcal{D}%
^{S}(\rho _{1}^{S}(\tau _{n}^{\min }), \rho _{2}^{S}(\tau _{n}^{\min }))=0.$
Therefore, Eq. (1) is reduced to $\mathcal{N}^{S}=\underset{\rho _{1,2}(0)}{%
\max }\sum_{n}\mathcal{D}^{S}\left( \rho _{1}^{S}\left( \tau _{n}^{\max
}\right) ,\rho _{2}^{S}\left( \tau _{n}^{\max }\right) \right) $ and the
maximum taken over all pair of initial states is equivalent to finding a
trace distance whose local maxima are larger than those of others. In what
follows, we adopt $\mathcal{D}_{\mathcal{N}}^{S}\left( \rho _{1}^{S}(t),\rho
_{2}^{S}(t)\right) $ for such a trace distance and
\begin{equation}
\mathcal{N}^{S}=\sum_{n}\mathcal{D}_{\mathcal{N}}^{S}\left( \rho
_{1}^{S}\left( \tau _{n}^{\max }\right) ,\rho _{2}^{S}\left( \tau _{n}^{\max
}\right) \right)
\end{equation}
for the non-Markovianity of a single qubit case, where the summation is over
all local maxima of $\mathcal{D}_{\mathcal{N}}^{S}\left( \rho
_{1}^{S}(t),\rho _{2}^{S}(t)\right) .$ We first prove the theorem below.

Theorem: \textit{There exists a maximum trace distance }$\left\vert
b(t)\right\vert $ \textit{at any instant time in Eq. (7) when }$\alpha =\mu
=1/2,$\textit{\ }$\left\vert \beta \right\vert =\left\vert \nu \right\vert $
$=1/2$ and $\left\vert \beta -\nu \right\vert =1$\textit{.}

Proof: Suppose $\mathcal{D}^{S}\left( \rho _{1}^{S}(t),\rho
_{2}^{S}(t)\right) =\left\vert b(t)\right\vert d(t)$ with $d(t)=\sqrt{%
\left\vert b(t)\right\vert ^{2}\left( \alpha -\mu \right) ^{2}+\left\vert
\beta -\nu \right\vert ^{2}}$ where $d(t)$ can be taken as the distance
between the points $P_{1}\left( \left\vert b(t)\right\vert \alpha ,\beta
\right) $ and $P_{2}\left( \left\vert b(t)\right\vert \mu ,\nu \right) ,$
and $P_{i}(x,y)$ $(i=1,2)$ denotes the points with $x\in \mathcal{\
\mathbb{R}
}$ and $y\in \mathcal{%
\mathbb{C}
}$. It is convenient to check that [$\left\vert b(t)\right\vert \alpha
-\left\vert b(t)\right\vert /2]^{2}/[\left\vert b(t)\right\vert
/2]^{2}+\left\vert \beta \right\vert ^{2}/\left( 1/2\right) ^{2}-1=4\left[
\left\vert \beta \right\vert ^{2}-\alpha (1-\alpha )\right] \leq 0,$ which
implies that the point $P_{1}\left( \left\vert b(t)\right\vert \alpha ,\beta
\right) $ or $P_{2}\left( \left\vert b(t)\right\vert \mu ,\nu \right) $ is
in (or on the circumference of) the ellipse
\begin{equation}
\lbrack x-\left\vert b(t)\right\vert /2]^{2}/[\left\vert b(t)\right\vert
/2]^{2}+\left\vert y\right\vert ^{2}/(1/2)^{2}=1,
\end{equation}%
with $x\in \mathcal{%
\mathbb{R}
}$ and $y\in \mathcal{%
\mathbb{C}
}$. We know that the maximum distance between the two points in an ellipse
is in between the two ends of the major axis. Since $\left\vert
b(t)\right\vert <1$ $(t>0),$ $d(t)$ reaches the maximum 1 only in the case
of $\left\vert b(t)\right\vert \alpha =$ $\left\vert b(t)\right\vert \mu
=\left\vert b(t)\right\vert /2,$ i.e., $\alpha =\mu =1/2$ and $\left\vert
\beta \right\vert =\left\vert \nu \right\vert $ $=1/2$ and $\left\vert \beta
-\nu \right\vert =1$. Consequently, $\mathcal{D}^{S}\left( \rho
_{1}^{S}(t),\rho _{2}^{S}(t)\right) $ will also reach the maximum $%
\left\vert b(t)\right\vert $. $\blacksquare $

It is easy to check that any pair of initial states satisfying the
conditions in above theorem definitely owns the same trace distance $%
\left\vert b(t)\right\vert $. Since the maximum of the trace at any instant
time is $\left\vert b(t)\right\vert $, for any pair of the initial states
not meeting the conditions in above theorem, the local maxima of $\left\vert
b(t)\right\vert $ should be never larger than those of the initial pairs
meeting the conditions. In what follows, we will employ $\mathcal{D}_{%
\mathcal{N}}^{S}=\left\vert b(t)\right\vert $ as the trace distance for
measuring non-Markovianity. As a result, the calculation of non-Markovianity
can be simplified to an easily computable expression
\begin{equation}
\mathcal{N}^{S}=\sum_{n}\left\vert b\left( \tau _{n}^{\max }\right)
\right\vert ,
\end{equation}%
with $\tau _{n}^{\max }$ the time point when $\left\vert b(t)\right\vert $
reaches the $n$th local maximum.

Straightforwardly, we can find the relationship between the population of a
single qubit initially in the excited state $\left\vert e\right\rangle $ and
the maximum trace distance $\mathcal{P}_{|e\rangle }=(\mathcal{D}_{\mathcal{%
N }}^{S})^{2},$ which yields
\begin{equation}
\mathcal{N}^{S}=\sum_{n}\sqrt{\mathcal{P}_{|e\rangle }(\tau _{n}^{\max })},
\end{equation}
from which the non-Markovianity of the qubit coupled to the reservoir could
be measured from the population of the upper level of the qubit. The
requirements for this implementation are (1) the bosonic reservoir is
initially in vacuum state and (2) the initial state of the qubit is prepared
in the upper level.

\begin{figure}[tbp]
\centering
\includegraphics[width=3.2in,bb=69pt 191pt 550pt 609pt]{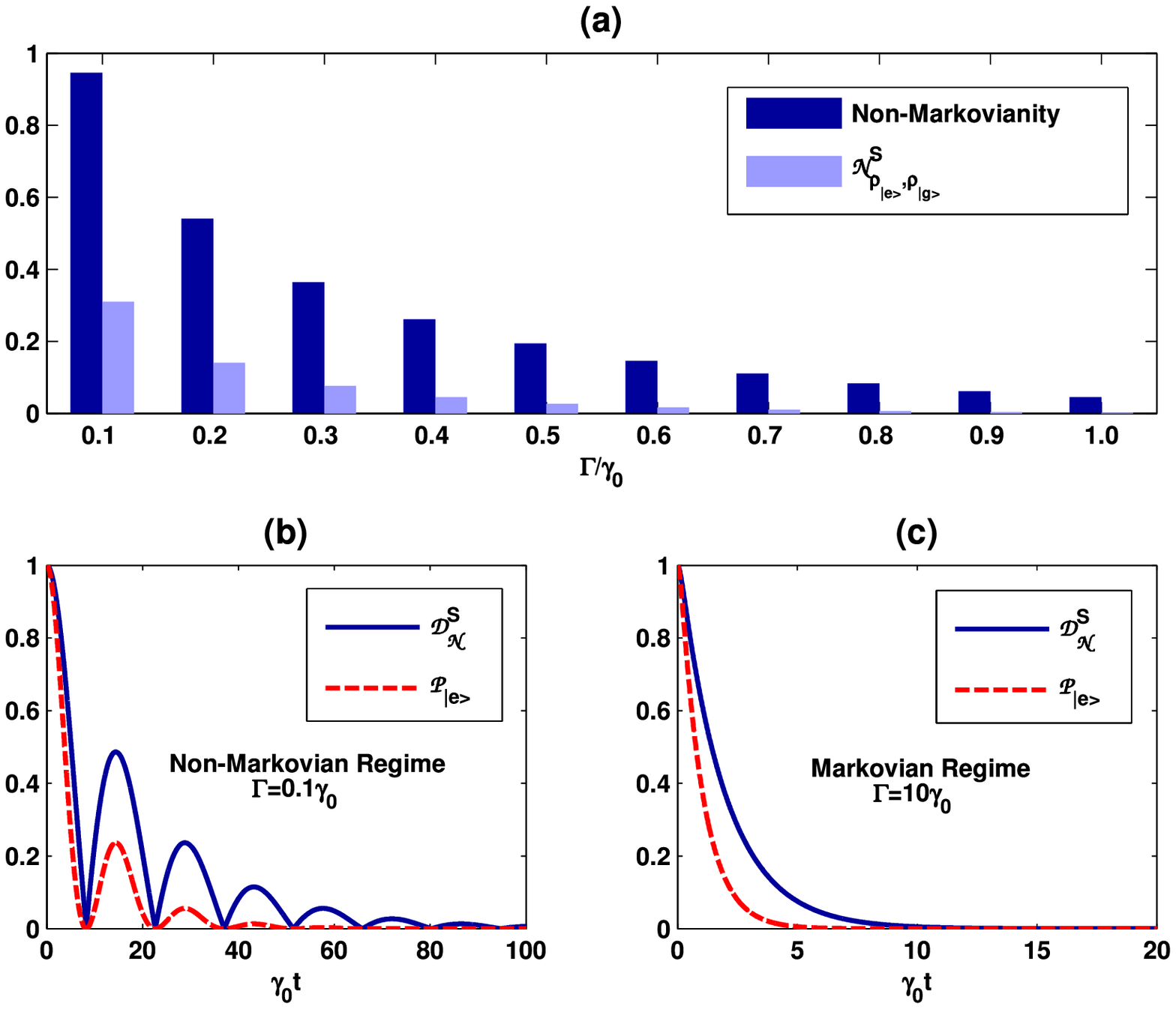}
\caption{(Color online) (a) Blue bars (black in the printed version): The
non-Markovianity of a single qubit coupled to a zero-temperature bosonic
reservoir with a Lorentzian spectrum. Light blue bars (gray in the printed
version): The total growth of trace distance with the initial pair of states
$\protect\rho _{\left\vert e\right\rangle }(0)=\left\vert e\right\rangle
\left\langle e\right\vert $ and $\protect\rho _{\left\vert g\right\rangle
}(0)=\left\vert g\right\rangle \left\langle g\right\vert .$ The maximum
trace distance $\mathcal{D}_{\mathcal{N}}^{S}\mathcal{\ }$(blue solid lines)
characterizes the decay process (red dashed lines) of a single qubit in (b)
non-Markovian regime (e.g., $\Gamma =0.1\protect\gamma _{0}$) and (c)
Markovian regime (e.g., $\Gamma =10\protect\gamma _{0}$) respectively.}
\label{fig1}
\end{figure}

Figure 1(a) shows a comparison between the non-Markovianity (the dark blue
bars) and $\mathcal{N}_{\rho _{\left\vert e\right\rangle ,}\rho _{\left\vert
g\right\rangle }}^{S}$ (the light blue bars) in the case of non-Markovian
regime ranging from $\Gamma =0.1\gamma _{0}$ to $\Gamma =\gamma _{0},$ where
$\mathcal{N}_{\rho _{\left\vert e\right\rangle ,}\rho _{\left\vert
g\right\rangle }}^{S}$ is measured by the trace distance $\mathcal{D}%
^{S}\left( \rho _{\left\vert e\right\rangle }(t),\rho _{\left\vert
g\right\rangle }(t)\right) =\left\vert b(t)\right\vert ^{2}$ with the
initial pair of states $\rho _{\left\vert e\right\rangle }(0)=\left\vert
e\right\rangle \left\langle e\right\vert $ ($\alpha =1,\beta =0$) and $\rho
_{\left\vert g\right\rangle }(0)=\left\vert g\right\rangle \left\langle
g\right\vert $ ($\mu =0,\nu =0$). Due to $\left\vert b(t)\right\vert <1$ and
$\left\vert b(t)\right\vert ^{2}<\left\vert b(t)\right\vert $, the light
blue bars are always shorter than the dark blue ones. In addition, it can be
seen that the indistinguishability of the states grows with the increase of
reservoir bandwidth $\Gamma $. This might be interpreted as the
non-Markovian character becoming less evident when the coupling between the
qubit and the reservoir decreases.

We have studied in Fig. 1(b) the population of the excited state $\mathcal{P}%
_{|e\rangle }$ and $\mathcal{D}_{\mathcal{N}}^{S}$ in the non-Markovian
regime (e.g., $\Gamma =0.1\gamma _{0}$), which shows the population $%
\mathcal{P}_{|e\rangle }$ reviving with the increase of $\mathcal{D}_{%
\mathcal{N}}^{S}$. This could be explained as the non-Markovian effect of
the bosonic reservoir: The process with $\mathcal{D}_{\mathcal{N}}^{S}$
being larger corresponds to the case that information lost by the qubit
flows back from the reservoir, increasing the distinguishability. So the
population revives for several times during this period. In contrast to the
non-Markovian regime, no revival of population occurs in the weak coupling
Markovian regime (e.g., $\Gamma =10\gamma _{0}$) as shown in Fig. 1(c),
because $\mathcal{D}_{\mathcal{N}}^{S}$ monotonously approaches zero with $%
b(t)\simeq \exp [-(\Gamma -\kappa )t/2]$. This implies no information
flowing back to the qubit.

The above analysis can be conveniently extended to other spectral densities,
other than Lorentzian spectral distribution, for non-Markovian
characterization of a qubit coupled to a bosonic reservoir \cite{density},
as long as the condition $\left\vert b(\tau _{n}^{\min })\right\vert =0$ ($%
n=1,2,3,\cdots $) is fulfilled \cite{Note}.

Let us take a brief look at the non-Markovianity of two identical
non-interacting qubits $A$ and $B$ locally interacting with two independent
zero-temperature bosonic reservoirs respectively. We noticed that taking the
maximum over any pair of initial states according to Eq. (1) is nearly
intractable in two-qubit case, although numerical simulation might been
employed for this job \cite{non-Markovianity}. However, since any growth of
the trace distance is a clear illustration of non-Markovian character, in
the following, instead of finding the maximum, we consider the trace
distance
\begin{equation}
\mathcal{D}^{T}\left( \rho _{\left\vert ++\right\rangle }(t),\rho
_{\left\vert --\right\rangle }(t)\right) =|b(t)|\sqrt{%
2-2|b(t)|^{2}+|b(t)|^{4}}
\end{equation}%
as the measurement of the non-Markovian character of the two qubits, where
the initial pair of states $\rho _{\left\vert ++\right\rangle
}(0)=\left\vert +\right\rangle _{A}\left\langle +\right\vert \otimes
\left\vert +\right\rangle _{B}\left\langle +\right\vert $ and $\rho
_{\left\vert --\right\rangle }(0)=\left\vert -\right\rangle _{A}\left\langle
-\right\vert \otimes \left\vert -\right\rangle _{B}\left\langle -\right\vert
$ with $\left\vert \pm \right\rangle _{i}=(\left\vert g\right\rangle _{i}\pm
\left\vert e\right\rangle _{i})/\sqrt{2}$ and $i=A,B$. Similar to the
single-qubit case, the trace distance approaches zeros, i.e., $\mathcal{D}%
^{T}\left( \rho _{\left\vert ++\right\rangle }(\tau _{n}^{\min }),\rho
_{\left\vert --\right\rangle }(\tau _{n}^{\min })\right) =0$ when $%
\left\vert b(\tau _{n}^{\min })\right\vert =0.$ Therefore, a lower bound of
non-Markovianity can be calculated by
\begin{equation}
\mathcal{N}_{\rho _{\left\vert ++\right\rangle },\rho _{\left\vert
--\right\rangle }}^{T}=\sum_{n}|b(\tau _{n}^{\max })|\sqrt{2-2|b(\tau
_{n}^{\max })|^{2}+|b(\tau _{n}^{\max })|^{4}},
\end{equation}%
or by the equivalent form
\begin{equation}
\mathcal{N}_{\rho _{\left\vert ++\right\rangle },\rho _{\left\vert
--\right\rangle }}^{T}=\sum_{n}\sqrt{2\mathcal{P}_{|e\rangle }(\tau
_{n}^{\max })-2\mathcal{P}_{|e\rangle }(\tau _{n}^{\max })^{2}+\mathcal{P}%
_{|e\rangle }(\tau _{n}^{\max })^{3}}.
\end{equation}%
Similar to the single-qubit case, the non-Markovian character lowers with
the increase of spectral width of the coupling, as shown in Fig. 2(a).

\begin{figure}[tbp]
\centering
\includegraphics[width=3.2in,bb=71pt 195pt 545pt 606pt]{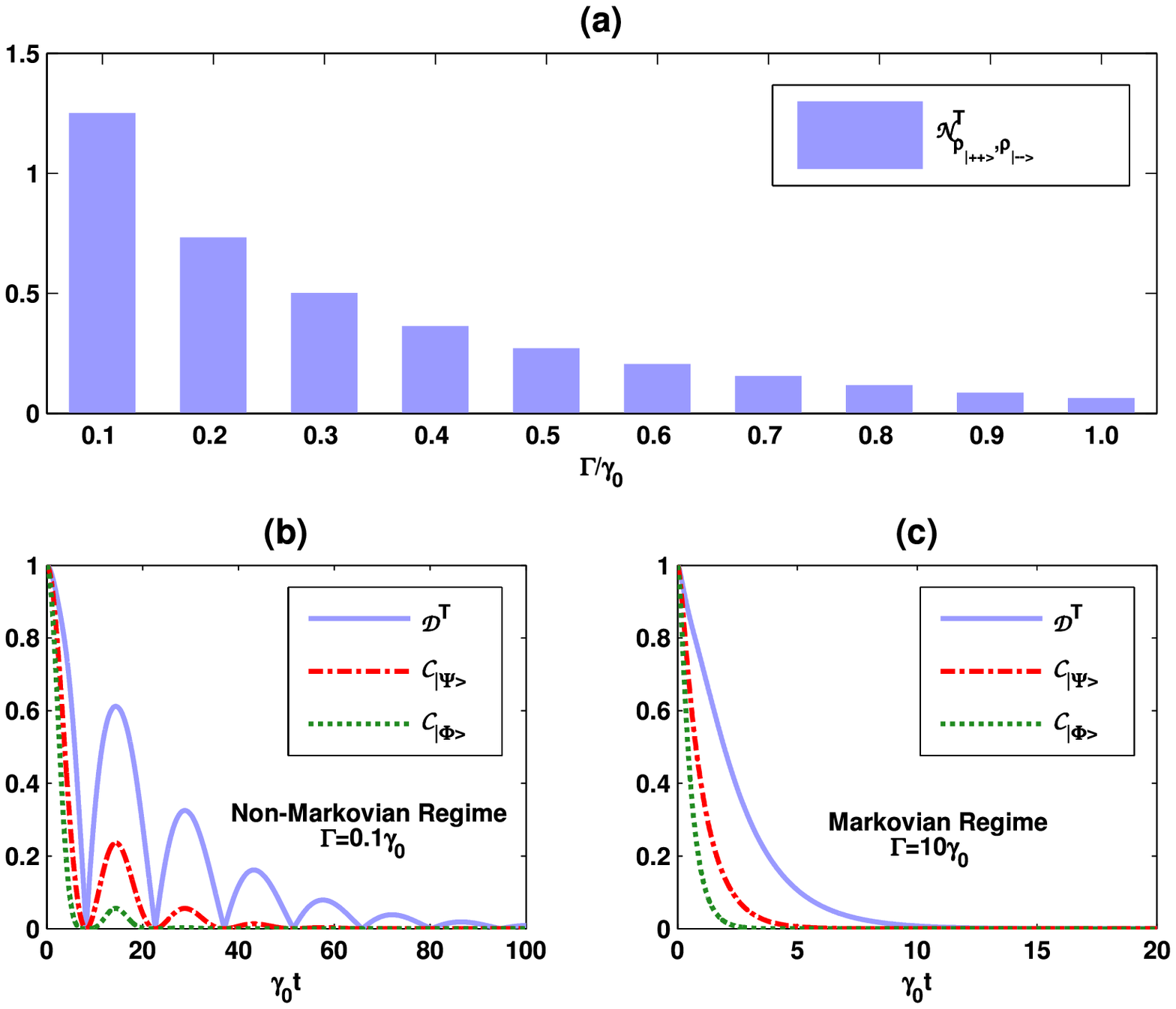}
\caption{(Color online) (a) A lower bound of non-Markovianity corresponding
to the trace distance $\mathcal{D}^{T}$ with initial pair of states $\protect%
\rho _{\left\vert ++\right\rangle }(0)=\left\vert +\right\rangle
_{A}\left\langle +\right\vert \otimes \left\vert +\right\rangle
_{B}\left\langle +\right\vert $ and $\protect\rho _{\left\vert
--\right\rangle }(0)=\left\vert -\right\rangle _{A}\left\langle -\right\vert
\otimes \left\vert -\right\rangle _{B}\left\langle -\right\vert $ ($%
\left\vert \pm \right\rangle _{i}=(\left\vert g\right\rangle _{i}\pm
\left\vert e\right\rangle _{i})/\protect\sqrt{2},i=A,B$). The trace distance
$\mathcal{D}^{T}$ (blue solid lines) characterizes the dynamics of two
non-interacting qubits coupling to two independent reservoirs in (b)
non-Markovian regime (e.g., $\Gamma =0.1\protect\gamma _{0})$ and (c)
Markovian regime (e.g., $\Gamma =10\protect\gamma _{0})$. While the red
dot-dashed lines and green dotted lines represent the concurrence of two
qubits initially prepared in Bell states $\left\vert \Psi \right\rangle
=(\left\vert ge\right\rangle +\left\vert eg\right\rangle )/\protect\sqrt{2}$
and $\left\vert \Phi \right\rangle =(\left\vert gg\right\rangle +\left\vert
ee\right\rangle )/\protect\sqrt{2}$, respectively.}
\label{fig2}
\end{figure}

Conventionally, the dynamics of two entangled qubits can be measured by
Wootters concurrence \cite{concurrence}. Using the method in Ref. \cite%
{two-qubit}, we obtain the concurrence of the two qubits as
\begin{eqnarray}
\mathcal{C}_{\left\vert \Psi \right\rangle } &=&\max \{0,\left\vert
b(t)\right\vert ^{2}\},  \notag \\
\mathcal{C}_{\left\vert \Phi \right\rangle } &=&\max \{0,\left\vert
b(t)\right\vert ^{4}\},
\end{eqnarray}%
when the initial states are prepared in Bell states $\left\vert \Psi
\right\rangle =(\left\vert ge\right\rangle +\left\vert eg\right\rangle )/%
\sqrt{2}$ and $\left\vert \Phi \right\rangle =(\left\vert gg\right\rangle
+\left\vert ee\right\rangle )/\sqrt{2}$, respectively \cite{two-qubit}.

The trace distance $\mathcal{D}^{T}$ (Eq. (13)) measuring a lower bound for
non-Markovianity can well characterize the non-Markovian behavior of the
dynamics of two qubits. In Fig. 2(b), the dynamics of the qubits initially
prepared in Bell states $\left\vert \Psi \right\rangle $ (red dot-dashed
line) and $\left\vert \Phi \right\rangle $ (green dotted line),
respectively, in non-Markovian regime (e.g., $\Gamma =0.1\gamma _{0})$ is
investigated. The entanglement measured by concurrence periodically vanishes
in accordance with the trace distance $\mathcal{D}^{T}$. In contrast, the
trace distance $\mathcal{D}^{T}$, in the Markovian regime (e.g., $\Gamma
=10\gamma _{0}$), asymptotically approaches zero, as shown in Fig. 2(c)
which implies that no entanglement revival exists in Markovian regime.

In practice, our scheme would be very preferable for experimental
implementation. According to Eqs. (12) and (15), we can study the
non-Markovian effect on coherence and entanglement of the qubits coupled to
bosonic reservoirs by directly measuring the population of a qubit without
resorting to tomographic reconstruction of the density matrix. Besides, our
proposal requires no specific information about the interaction between the
qubit and the reservoir. These favorable features make our scheme feasible
in experiments under real environments, e.g., using two-level atoms confined
in optical microcavities \cite{OPC} or under simulated reservoirs \cite{QS},
e.g., using a spin-reservoir model with spectral densities ranging from sub
ohmic to super ohmic cases simulated by trapped ions \cite{QS-ion}.

To summarize, we have presented a simple method for measuring the
non-Markovian character of the qubits coupled to bosonic reservoirs. We
believe that this easily operated measure for non-Markovianity would be very
useful for further understanding non-Markovian behavior and also for
experimental exploration, particularly in a realistic experimental
situations without knowing much about the interaction between a qubit and
the environment.

Z.Y.X. is grateful to Hua Wei for his warmhearted help. This work is
supported by National Natural Science Foundation of China (NNSFC) under
Grant No. 10774163.


\begin{thebibliography}{99}
\bibitem{open book} H.-P. Breuer and F. Petruccione, \textit{The Theory of
Open Quantum Systems} (Oxford University Press, Oxford, 2007).

\bibitem{open book2} C. W. Gardiner and P. Zoller, \textit{Quantum Noise}
(Springer-Verlag, Berlin, 2004); U. Weiss, \textit{Quantum Dissipative
Systems} (World Scientific Publishing, Singapore, 2008).

\bibitem{QIP Book} M. A. Nielsen and I. L. Chuang, \textit{Quantum
Computation and Quantum Information} (Cambridge University Press, Cambridge,
England, 2000).

\bibitem{exp-molecular} P. Hamm \textit{et al}., Phys. Rev. Lett. \textbf{81}%
, 5326 (1998); V. O. Lorenz and S. T. Cundiff, Phys. Rev. Lett. \textbf{95},
163601 (2005).

\bibitem{exp-High-Q} F. Dublin \textit{et al}., Phys. Rev. Lett. \textbf{98}%
, 183003 (2007).

\bibitem{exp-solid} C. W. Lai \textit{et al}., Phys. Rev. Lett. \textbf{96},
167403 (2006); D. Mogilevtsev \textit{et al}., Phys. Rev. Lett. \textbf{100}%
, 017401 (2008); C. Galland \textit{et al}., Phys. Rev. Lett. \textbf{101},
067402 (2008).

\bibitem{quantum trajectory} W. T. Strunz \textit{et al}., Phys. Rev. Lett.
\textbf{82}, 1801 (1999); A. A. Budini, Phys. Rev. A \textbf{63}, 012106
(2000); H.-P. Breuer, Phys. Rev. A \textbf{70}, 012106 (2004); A. A. Budini,
Phys. Rev. A \textbf{74}, 053815 (2006); A. Bassi and L. Ferialdi, Phys.
Rev. Lett. \textbf{103}, 050403 (2009).

\bibitem{pseudomode} B. M. Garraway, Phys. Rev. A \textbf{55}, 2290 (1997);
L. Mazzola \textit{et al}., Phys. Rev. A \textbf{80}, 012104 (2009).

\bibitem{NMQJ} J. Piilo \textit{et al}., Phys. Rev. Lett. \textbf{100},
180402 (2008); J. Piilo \textit{et al}., Phys. Rev. A \textbf{79}, 062112
(2009); H.-P. Breuer and J. Piilo, Europhys. Lett. \textbf{85}, 50004 (2009).

\bibitem{semi-MP} H.-P. Breuer and B. Vacchini, Phys. Rev. Lett. \textbf{101}%
, 140402 (2008); H.-P. Breuer and B. Vacchini, Phys. Rev. E \textbf{79},
041147 (2009).

\bibitem{M} M. M. Wolf \textit{et al}., Phys. Rev. Lett. \textbf{101},
150402 (2008).

\bibitem{non-Markovianity} H.-P. Breuer \textit{et al}., Phys. Rev. Lett.
\textbf{103}, 210401 (2009).

\bibitem{nonM2} \'{A}. Rivas \textit{et al}., arXiv:0911.4270.

\bibitem{nonM3} X.-M. Lu \textit{et al}., arXiv:0912.0587.

\bibitem{Q-optics} M. O. Scully and M. S. Zubairy, \textit{Quantum Optics}
(Cambridge University Press, New York, 1997); D. F. Walls and G. J. Milburn,
\textit{Quantum Optics} (Springer Verlag, Berlin, 2008).

\bibitem{density} A. J. Leggett \textit{et al}., Rev. Mod. Phys. \textbf{59}%
, 1 (1987).

\bibitem{Note} However, if $\left\vert b(\tau _{n}^{\min })\right\vert \neq
0,$ we may only acquire a lower bound of the non-Markovianity and Eqs. (11)
and (12) will, therefore, be modified to $\mathcal{N}^{S}=\sum_{n}(\left%
\vert b\left( \tau _{n}^{\max }\right) \right\vert -\left\vert b\left( \tau
_{n}^{\min }\right) \right\vert )$ and $\mathcal{N}^{S}=\sum_{n}(\sqrt{%
\mathcal{P}_{|e\rangle }(\tau _{n}^{\max })}-\sqrt{\mathcal{P}_{|e\rangle
}(\tau _{n}^{\min })})$ respectively.

\bibitem{concurrence} W. K. Wootters, Phys. Rev. Lett. \textbf{80}, 2245
(1998).

\bibitem{two-qubit} B. Bellomo \textit{et al}., Phys. Rev. Lett. \textbf{99}%
, 160502 (2007).

\bibitem{OPC} K. J. Vahala, Nature (London) \textbf{424}, 839 (2003).

\bibitem{QS} I. Buluta and F. Nori, Science \textbf{326}, 108 (2009).

\bibitem{QS-ion} D. Porras \textit{et al}., Phys. Rev. A \textbf{78},
010101(R) (2008).
\end{thebibliography}
\end{document}